\documentclass[CEJP,DVI]{cej} % use DVI command to enable LaTeX driver
\usepackage{layout}
\usepackage{amsmath}
\usepackage{textcomp}
\usepackage{graphicx}
\title{Electron-phonon interaction in nanodevices}

\articletype{Research Article}

\author{Karel Kr\'al\inst{1}\email{kral@fzu.cz}
}

\institute{
     \inst{1} Institute of Physics, Academy of Sciences of Czech Republic,\\
     Na Slovance 2, 18221 Prague 8, Czech Republic
          }

\abstract{The effect of the up-conversion of the electronic energy
level occupation was earlier interpreted as an implication of the multiple scattering of the charge carriers on the longitudinal optical phonons of the lattice vibrations in a small system like a quantum dot. In this work we study the influence of this effect on the electronic motion in a
nanotransistor represented by a quantum dot connected to two
electric wires and a gate electrode. We show that in an asymmetric
nanotransistor the up-conversion effect gives rise to a
spontaneous current between the source and the drain, or to an appearance of a spontaneous voltage between these electrodes. The effect will be
studied basing on the well known Datta's Toy Model of the
theoretical description of the nanotransistor and on additional
kinetic equations giving the multiple scattering of electrons in the
quantum dot, in the self-consistent Born approximation to the
electronic self-energy. We shall also briefly discuss the relation of
this theoretical result to existing experiments on current-voltage
characteristics in gated nanostructures.}

\begin{document}
\maketitle

%% ###################################################################

\section{Introduction}
%*********************

Semiconductor quantum dots connected to electric contacts are a
prototype of electronic elements interesting for the future
nanoelectronics. The bound state energy level separations of the
charge carriers in these structures are close to the excitation
energies of the lattice vibrations, like the longitudinal optical (LO) phonons. This property, together with the quasi-zero-dimensional character of the dots, which may emphasize the role of the multiple-scattering  of the charge carriers  moving along the atomic structure, leads to an
enhanced role of the optical phonons in the quantum dot structures.

The electrons of the zero-dimensional nanostructures are coupled to
the bath via atomic vibrations of the zero-dimensional nanostructure
\cite{Nozik2001} or, as it is in the case of molecules via the molecular
collisions \cite{Tyakht}. The electron interaction with the
LO phonons has a significant impact on the electronic energy
relaxation and up-conversion and also on the line shape of the
optical emission in quantum dots \cite{Nozik2001}-\cite{Boxberg}.

The multiple scattering of electrons on the longitudinal optical
(LO) phonons in quantum dots, when included in the electronic
quantum kinetic equation in the self-consistent Born approximation
to the electronic self-energy, gives not only the fast electronic
energy relaxation in these nanostructures, but also the effect of
the up-conversion of electronic level occupation\cite{SurfSci}.
The same mechanism is likely to occur in the open systems of
nanostructures, like in those of a quantum dot connected to metallic
electrodes.

The effect of the electronic occupation up-conversion can be seen as
an implication of the nonadiabaticity of the electron-phonon motion.
Let us remind that the electron moving in a quantum dot is forced to
make multiple scattering acts on the LO phonons, because of the
space restriction of the electronic motion inside the quantum dot.
This can be expected to lead to the presence of multiphonon states
in the dot. These states may be regarded as being similar to coherent
photon states, which again may resemble classical waves of electric
field in the coherent laser pulse. As such the multiphonon lattice
oscillations influence the electronic motion in the dot as a time
dependent potential. The effective Hamiltonian of the electronic
motion in a quantum dot can be then regarded as explicitly time dependent. From the point of view of the electronic energy the system can
be then viewed as non-conservative.

The electronic energy level occupation in quantum dots, influenced
by the up-conversion of the electronic energy levels, discussed
recently in individual quantum dots\cite{SurfSci}, can be important also
for the electric current through an open quantum dot, like that
connected to two metallic wires\cite{PaulssonZahidDatta}. In such a
setup the quantum dot may be also connected to a third contact,
called a gate, as it is shown in Fig.\ \ref{transistor2}. The
connection to the wires means that electronic subsystem of the dot
is coupled to the reservoirs of electrons in the two electric
contacts $L$ and $R$ via an electron tunneling coupling. The
coupling of the quantum dot to the contacts $L$, $R$ was shown
earlier to influence considerably the key properties of the
nanotransistor. This is clarified in the remarkable approach of the
so called Toy Model of Paulsson, Zahid and Datta
\cite{PaulssonZahidDatta}. This Toy Model will be applied here.

The fast electronic relaxation, particularly the up-conversion, and
the optical emission properties of quantum dots \cite{Nozik2001},
have been paid attention in connection with the effect of the
optical emission of the quantum dot lasers lasing from higher
excited states \cite{Rafailov,SurfSci}. The problem of the
up-conversion has also been studied under a non-integer
charging of the quantum dot by
electrons \cite{SurfSci,Vancouver2007}.

In this work we will show the influence of the electron-LO-phonon
scattering on the current-voltage characteristics of the
nanotransistor. We shall also show a possible connection between
the present theoretical results some recent experiments on open
nanosystems.

\section{Nanotransistor}
%*************************

\begin{figure}[th]
%\begin{center}
\includegraphics[width=5cm]{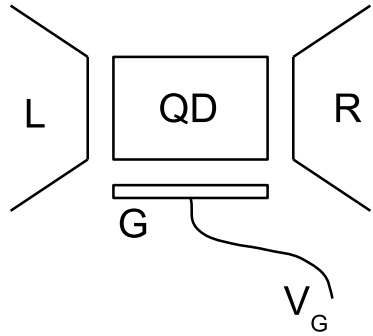}
\caption{The nanotransistor scheme is shown consisting of the
quantum dot (QD) connected to the metallic contacts $L$ and  $R$.
The letter $G$ denotes the gate electrode connected to the potential
$V_G$.\label{transistor2}}
%\end{center}
\end{figure}

A semiconductor quantum dot, in the setup of the nanotransistor, as
shown in Fig.\ \ref{transistor2}, determines its principal
properties by the position of the single-electron energy levels
\cite{PaulssonZahidDatta}. When a voltage is attached to the $L$ and
$R$ electrodes then the current through the device depends on the
electrochemical potentials in the two wires compared to the position
of the quantum dot single electron energies. For simplicity,
let us assume that the latter energies are suitably positioned along the electronic energy scale by the
effect of the gate electrode $G$. The resulting current flowing
through such a device depends also on the intensity of the
electronic tunneling interaction between the quantum dot
and the wires $L$ and $R$. This elastic tunneling interaction
transfers the carriers between the electrodes and the dot.
In the original version of the Toy Model the electrons are not
exchanged between the electronic states inside the quantum dot by
any intra-dot mechanism. The possibility of the electronic
inter-level transfer due to the LO phonons will be considered here.
We will see that the electronic up-conversion is then manifested in this device by an
appearance of a spontaneous voltage between the electrodes $L$
and $R$ in a nanotransistor in which the quantum dot is coupled
to the wires $L$ and $R$ in an asymmetric way.

\section{Electron kinetic equation}
%************************************

The basic part of the quantum kinetic equation giving the electric current through the nanotransistor will be constructed using the Toy Model \cite{PaulssonZahidDatta}. The additional mechanism, the electronic redistribution mechanism due to phonons, will be included as a separate procedure added to the Toy Model.

Besides the up-conversion mechanism we shall also include the
recently studied artificial effect of the overheating of the system
of LO phonons \cite{KralCJP2006,KralLinIJMPB}. This effect has been
ascribed to a shortcoming of the optical phonon kinetic equation.
The overheating effect can be eliminated \cite{KralCJP2006,
KralLinIJMPB} to a reasonable extent with help of a procedure built
on the Lang-Firsov canonical transformation. In the present work we
shall use this elimination process throughout the calculations.

Two nondegenerate electronic eigenstates \cite{SurfSci} in the quantum dot will
be denoted by indexes $A$ and $B$. These two states have their
respective unperturbed electronic energy levels $E_A$ and $E_B$, $E_B>E_A$. The electronic correlation interaction and spin will be neglected.
The electrons with different spins will be treated as independent. The Hamiltonian of the nanotransistor, corresponding to the scheme of the Fig.\
\ref{transistor2}, consists of the free electrons in the two
unperturbed electronic levels of the dot and the system of bulk LO
phonon modes of the dot atomic structure vibrations. The electrons
in each one of the energy levels are coupled to the contacts by a
tunneling mechanism. The electronic motion in a quantum dot will
be assumed to be influenced weakly by the electronic tunneling
coupling to the contacts $L$ and $R$. We shall assume that the
broadening of the single-electron energy levels in the quantum
dot will be of the order of one meV, which is then much smaller
than the expected electronic energy level separation in an
isolated quantum dot. The energy level separation in the dots having
the lateral size of tens of nanometers would be several tens of meV.
Let us remind that the magnitude of the longitudinal optical (LO)
phonon energy in GaAs crystalline material is of the same order of
magnitude (36.2 meV). Along with the LO phonons also other lattice
vibration modes may influence the electronic motion. We shall
neglect these degrees of freedom in the present work\cite{SurfSci}.

We shall use a parametrization of electronic tunneling coupling
between the dot and the wires in the form of level broadening
parameter $\Gamma$ \cite{PaulssonZahidDatta}. This parameter
$\Gamma$  is assumed to be independent of the intradot
electron-phonon coupling.

The effect of the third gate electrode will be expressed simply
by adding a suitable energy $E_G$ to all the single electron
unperturbed energies.

In the Toy Model kinetics the operator of the electronic transfer
between the quantum dot and the electrodes is a basic component of
the irreversible transfer of the charge carriers. Referring the
reader to the original work \cite{PaulssonZahidDatta} for details,
the rate of transfer of the electrons can be written down in a
straightforward way. For example, the electric current $I_{LA}$ from
the left wire  into the energy level $A$  of the quantum dot is
$I_{LA}=(-e\Gamma_{LA}/h)(N_{LA}-N_A)$, where $-e$, $e>0$, is
electronic charge. In the latter formula $N_A$ is the electronic
occupation of the energy level $A$. The quantity $N_{LA}$ is the
"target" occupation of the level $A$. It is such an occupation,
which the energy level $A$ tries to achieve in order to get into the
thermodynamic equilibrium with the left electrode ($L$). The
parameter $\Gamma_{LA}$  means the $A$-level broadening due to
virtual electronic transitions provided by the tunneling
transfer Hamiltonian \cite{PaulssonZahidDatta}.

\begin{figure}[th]
%\begin{center}
\includegraphics[width=7cm]{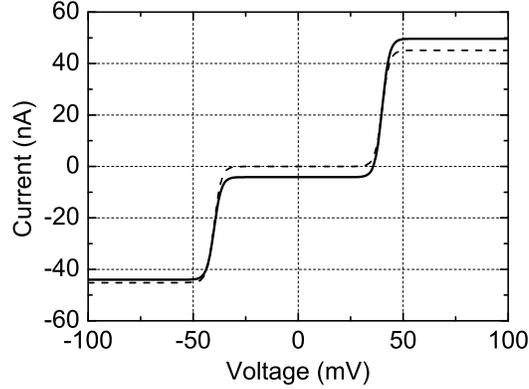}
\caption{Dependence of the current through the device as a function
of the applied voltage. The energies of the quantum dot are $E_A=0$
meV and  $E_B=40$ meV. The chemical potential of the unperturbed
wires is set to 20 meV. The current voltage characteristics for the
case without the effect of the electronic up-conversion is shown as
the dotted line. The electronic up-conversion is included with using the empirical analytical formula, see text.  The current with the up-conversion effect is shown
by the full line.\label{current1-3}}
%\end{center}
\end{figure}

The phonons generally play an important role in the low-dimensional
nanostructures \cite{Heeger1988}. We shall extend the Toy Model to
include also the electron-LO-phonon coupling. Let us consider a state when the level $A$ is occupied by $N_A$  electrons and the level $B$  is
occupied by $N_B$ electrons, the total occupation being $N_{tot}=N_A+N_B$. Referring the reader to the paper \cite{Vancouver2007} for the details, we just remind here that a simple approximate analytical formula can be found for  the irreversible process of the electronic relaxation and up-conversion between the two electronic states within the dot, which depends on the occupations $N_B$  and $N_{tot}$ and on other parameters characterizing the quantum dot. This formula was determined in an empirical way, for an isolated quantum dot without the attached wires \cite{Vancouver2007} or a gate. In other words, this formula does not contain any influence of the electronic coupling to the wires on the intra-dot inter-level relaxation. The analytical formula was established for the special case of the quantum dot used in this work, namely the quantum dot with the cubic shape, with the infinitely deep potential, having two
bound states. The lateral size of the dot is chosen to be 18 nm.
At this size the ratio between the electronic energy level separation
and the longitudinal optical phonon energy is about 1.44. The empirical analytical formula has been determined at the temperature of 10 K. We shall start the numerical computations with the electron-LO-phonon scattering included in the form of the analytical formula of the reference \cite{Vancouver2007}. The material constants will be those of the bulk GaAs crystal. Therefore, the numerical calculations will be first done for this particular situation in the quantum dot. Let us remark that using the analytical formula for the
electron-phonon relaxation makes the numerical computations
of the current-voltage characteristics fast. This approach
has also shortcomings, for example  the analytical formula
does not contain the dependence of the interlevel relaxation
rate on the lattice temperature \cite{Vancouver2007}.

Let us also remind that the above given relaxation rate formula was computed in an approximation which corresponds to the effect of the so called phonon overheating suppressed.  The reader is referred to the references \cite{KralCJP2006, KralLinIJMPB} for the details about the overheating problem.

As a second step in the numerical evaluation of the current-voltage characteristics of the nanotransistor,  we will compute the electronic relaxation rate between the electronic levels directly for the given state of the parameters of the dot without finding first an analytical formula. The characteristics of the quantum dot itself remain without changes. The intradot scattering on the LO phonons will be calculated directly using the relaxation rate equations \cite{SurfSci}.
In both ways of the numerical treatments of the electron-phonon interaction, the irreversible effect of the electron-phonon scattering \cite{KralKhas1998} is based on the self-consistent Born approximation to the electronic self-energy in the nonequilibrium Green's functions method of the quantum kinetic equations. This degree of approximation to the electronic self-energy is in a conceptual agreement with the recent statement \cite{Kapon2007}  that the electronic motion theory in quantum dots is out of reach of perturbative approaches. The common Kadanoff-Baym ansatz was applied to the quantum kinetics of the up-conversion and the instant collision approximation was used.

\section{Current-voltage characteristics of the nanotransistor}
%**************************************************************

The kinetic equations determine the electric current through the
device and the electronic level occupation at the given voltage
applied to the electrodes $L$  and  $R$. In order to simplify the argumentation we will assume now that a suitable gate potential is applied to help us to make the current voltage characteristics, without the phonon effect, to be symmetric. For this purpose we assume in particular that the gate potential is set such that at zero applied voltage, and with the phonons influence switched off, the chemical potential of the contacts $L$ and $R$ is placed into the middle of the gap between the energies $E_A$ and $E_B$. In the electron kinetic equations the energy levels of the quantum dot belong to input parameters, together with the applied voltage, the chemical potentials of the electrodes and the temperature. With these parameters the empirical formula for the rate of the electronic relaxation and the up-conversion are obtained \cite{Vancouver2007}.

An important point of this work is an asymmetry of the contacts. In the course of the calculations we assume an asymmetry of the parameters  determining the tunneling between the dot and the wires $L$ and $R$. We have the four tunneling parameters in the model of the device, namely, the parameters $\Gamma_{LA}$ and $\Gamma_{LB}$ are the tunneling coupling parameters between the left wire and the states $A$ and $B$, while $\Gamma_{AR}$ and $\Gamma_{AR}$ and $\Gamma_{BR}$ are the electron tunneling parameters between the respective states $A$ and $B$ and the wire $R$. We choose $\Gamma_{LB}=\Gamma_{LA}=\Gamma_{AR}=$1 meV, while $\Gamma_{BR}=$2 meV. With these values the Fig.\ \ref{current1-3} shows that the electronic up-conversion leads to a nonzero electric current of about -4 nA at zero voltage applied. We see that the electric current of -4 nA flows to the left hand side

\begin{figure}[th]
%\begin{center}
\includegraphics[width=7cm]{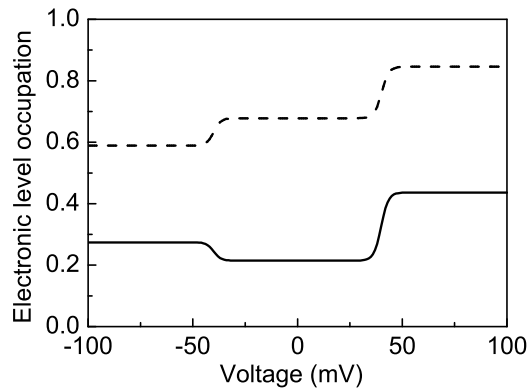}
\caption{The occupation of the electronic states  $A$ (dashed) and
$B$ (full) calculated at temperature 10 K and with the up-conversion
included. The other parameters correspond to the full line in the
Fig.\ \ref{current1-3}.\label{obsazeni01}}
%\end{center}
\end{figure}

even at a certain range of the voltage values, including a range of
positive values of the voltage. In other words, at these values of
the voltage the electric current flows in the direction from the
quantum dot to the electrode $L$, while the electrons themselves
flow from the quantum dot to the right hand side electrode $R$. This happens because
after being up-converted to the energy level $B$ the electrons find
it easier to go to the electrode $R$, due to the asymmetry in the
choice of the parameters $\Gamma$. The electrons at the energy
level $B$ can be exchanged between the quantum dot and the wire $R$
in a more easy way because of the increased value of the parameter
$\Gamma_{BR}$. The origin of the effect of the nonzero current can
thus be explained as an implication of the asymmetry of the electric
contacts in a cooperation with the non-adiabatic effect of the up-conversion of the
electrons to the energy level $B$. We have to keep in mind that we are obtaining this effect upon using the approximations specified above. Namely, so far we treat the electron-phonon scattering as a process independent of the electronic contact-to-quantum dot tunneling mechanism.

\begin{figure}[th]
%\begin{center}
\includegraphics[width=7cm]{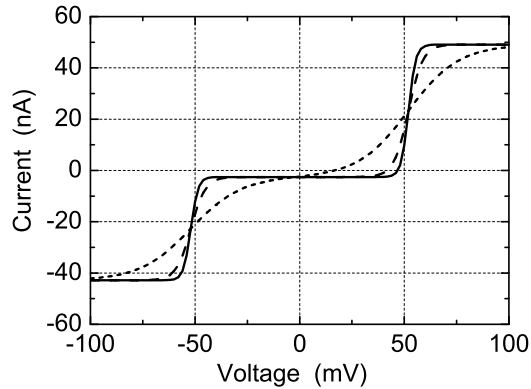}
\caption{Current-voltage characteristics of the nanotransistor including the up-conversion and the asymmetry of the contacts.  The calculation is improved with respect to that using the analytical current-voltage characteristics (see text). The electronic relaxation due to LO phonons is calculated from the relaxation rate formula. This formula is generalized to include the electronic tunneling to and from the wires $L$ and $R$. The electronic  part of the kinetic equation is given by the Toy
Model. Temperatures of the lattice: 10 K – full curve, 20 K – long dash, 70 K – short dash. Material parameters of bulk GaAs are used.    \label{proud102070-2}}
%\end{center}
\end{figure}

It is remarkable to see that the flow of the electrons to the right of the device becomes stopped at a certain value of the attached voltage, which is seen in the Fig.\ \ref{current1-3} as equal to about 35 mV. In practical cases this would mean that when the device is left alone with the contacts $L$  and $R$ disconnected from any other body and from one another, the electrons would flow to the right-hand-side electrode until the accumulated charge, with the given capacity of the device, create  a potential at the contacts which corresponds to the critical voltage of the 35 mV. In the real case, one has to expect that such a device would quickly capture an amount of compensating charges from its neighborhood and became electrostatically neutralized via parasitic currents. Transient properties of the presently considered device then could deserve an attention. In the transient processes the spontaneous creation of the voltage at the contacts might be more easily observable than in the course of a steady state process.

At some situations the quantity of the electrons at a single quantum dot may become rather low and the reduced relaxation rate of the electronic relaxation process due to the multiple electron-LO-phonon scattering may decrease considerably \cite{KralBoston2007,KralICTON2006,Vancouver2007}. Other relaxation mechanisms, like electron-acoustic phonon inelastic scattering may then become relatively important, the situation requiring a separate attention. From this reason, in Fig.\ \ref{obsazeni01} we verify that the total electronic occupation of our presently considered device remains well within the order of unity.

\section{Improvement of the model calculation}
%********************************************************

As a next step the calculation of the electric current through the nanotransistor is improved in two ways. First, the electronic relaxation due to the electron-phonon interaction is computed directly from the formula of the relaxation rate\cite{KralKhas1998}, instead of the use of the relaxation rate in the form of an approximate analytical formula \cite{Vancouver2007}. Second, the calculation of the relaxation rate \cite{KralKhas1998} is generalized to include also the tunneling of the quantum dot electrons to the electrodes $L$ and $R$ and back. This generalization has been performed at the level of assuming that the unperturbed electron energies $E_A$ and $E_B$ have an imaginary part corresponding to the magnitude of the tunneling coupling parameters $\Gamma$. In Fig.\ \ref{proud102070-2} we see the resulting current-voltage dependence.

The Fig. \ref{proud102070-2} shows that the calculations with using the analytical formula of the electronic relaxation rate, and those with the direct evaluation of the relaxation rate, give comparable results. Namely, the current through the nanotransistor shows again the tendency to display  the spontaneous electric current or voltage. Similarly as in the case of the use of the relaxation rate analytical formula Fig.\ \ref{current1-3}, we observe the spontaneous electric current of several  nano-Amperes at the zero applied voltage to the nanotransistor. Let us remark at this point that this value nevertheless depends on our knowledge of the magnitude of the parameters $\Gamma$.

With increasing the lattice temperature the original step-like form of the current-voltage characteristics becomes rather smeared already at the temperature of 70 K, see  Fig.\ \ref{proud102070-2}. This effect is ascribed to the increase of the  thermal occupation of the excited electronic state $B$.

\section{Remarks on experiments}
%********************************

Let us make a short remark on the recent experiments which can be supposedly related to the effect of the spontaneous current and voltage.

A large effort has continued recently in the field of the organic conducting materials. Among them the DNA molecule belongs to those subjects which displayed in the recent past remarkable and sometimes controversial properties of the electric conduction. One of the most attractive measurements are those in which the conductivity of individual organic molecules are measured. Short segments of DNA molecule have been measured with the help of AFM experimental setup \cite{Ullien}. In this setup the gate electrode $G$ is not included. The  measurements presented in the paper \cite{Ullien}, to which the reader is referred for details, display features  interesting from the point of view of the current-voltage characteristics  and the spontaneous current appearance, to the analysis of which the attention is paid in the present work. The short segments of DNA are complex enough for a theoretical analysis \cite{Enders}. From this reason the analysis of the DNA conduction may need a separate attention.

Another interesting experiment has been done on a GaAs nanotransistor \cite{Horsell}. The authors detect a spontaneous current in the circuit made of a shunted nanotransistor. The authors interpret the effect as due to the asymmetry of the nanodevice and the charge carriers thermally promoted to higher energy states in the active region of the nanotransistor. A detailed analysis of the observed electric current data would be desirable for  to obtain an estimate about to which extent  the electronic up-conversion based on the multi-LO-phonon scattering of electrons can participate on the  signal detected in the work \cite{Horsell}.

\section{Conclusions}
%*********************

In this paper we have used a simple model of the nanotransistor
based on the well-known Toy Model. The model has been extended to
include also the effect of the multiple scattering of electrons on
the optical phonons. The quantum kinetic equations based on the
self-consistent Born approximation to the electronic self-energy and
on the instant collision approximation, with the application of the
standard Kadanoff-Baym ansatz, have been used. The effect of the
electronic up-conversion, being an implication of the above
approximations, has been augmented to the Toy Model as an
independent process. The up-conversion has been treated within such
modification of the theory, in which the artificial effect of the
phonon overheating has been partly eliminated. The numerical results
based on these theoretical resources show an ability of the
asymmetric device to generate electric current, or to generate a
spontaneous voltage between the electric contacts. These effects are
attributed to the cooperation of the asymmetry of the electric
contacts in the small two-level device and the multiple scattering of electrons on optical phonons.
The experiments seem to provide interesting effects measured recently,   on both organic and inorganic materials.

\acknowledgements{This work was supported by the projects ME-866 of
Ministry of Education and by the project AVOZ10100520.}

\end{document}